\documentclass[preprintnumbers,prd,twocolumn,showpacs,floatfix,preprintnumbers,letterpaper,amsmath,amssymb,nofootinbib,superscriptaddress]{revtex4}
\usepackage{graphicx}
\usepackage{epsfig}
\usepackage{bm}
\usepackage{amsfonts}
\usepackage{slashed}
\usepackage{latexsym,float,url}

\def\bx{{\bf x}}
\def\bk{{\bf k}}
\def\tpk{{\tilde{\phi}_k}}

\def\be{\begin{equation}}
\def\ee{\end{equation}}
\def\ba{\begin{eqnarray}}
\def\ea{\end{eqnarray}}
\renewcommand{\(}{\left(}
\renewcommand{\)}{\right)}
\renewcommand{\[}{\left[}
\renewcommand{\]}{\right]}

\begin{document}

\title{Cosmological perturbations in $f(T)$ gravity}

\author{Shih-Hung Chen} \email{schen102@asu.edu}
 \affiliation{Department of Physics and School of Earth and Space
Exploration, Arizona State University, Tempe, AZ 85287-1404}

\author{James~B.~Dent} \email{jbdent@asu.edu}
\affiliation{Department of Physics and School of Earth and Space
Exploration, Arizona State University, Tempe, AZ 85287-1404}

\author{Sourish Dutta}
\email{sourish.d@gmail.com} \affiliation{Department of Physics and
Astronomy, Vanderbilt University, Nashville, TN  ~~37235}

\author{Emmanuel N. Saridakis}
\email{msaridak@phys.uoa.gr}
 \affiliation{College of Mathematics
and Physics,\\ Chongqing University of Posts and
Telecommunications Chongqing 400065, P.R. China }

\begin{abstract}
We investigate the cosmological perturbations in $f(T)$ gravity.
Examining the pure gravitational perturbations in the scalar
sector using a diagonal vierbien, we extract the corresponding dispersion relation, which
provides a constraint on the $f(T)$ ansatzes that lead to a theory
free of instabilities. Additionally, upon inclusion of the matter
perturbations, we derive the fully perturbed equations of motion,
and we study the growth of matter overdensities. We show that
$f(T)$ gravity with $f(T)$ constant coincides with General
Relativity, both at the background as well as at the first-order
perturbation level. Applying our formalism to the power-law model
we find that on large subhorizon scales ($\mathcal{O}$(100 Mpc) or
larger), the evolution of matter overdensity will differ from
$\Lambda$CDM cosmology.  Finally, examining the linear
perturbations of the vector and tensor sectors, we find that (for the standard choice of vierbein)
$f(T)$ gravity is free of massive gravitons.
\end{abstract}

 \pacs{98.80.-k, 04.50.Kd }

\maketitle

\section{Introduction}

Cosmological observations of the last decade \cite{c1} indicate
that the observable universe experiences an accelerated expansion.
Although the simplest way to explain this behavior is the
consideration of a cosmological constant \cite{c7}, the
difficulties associated with the fine-tuning problem have led some
authors to postulate more radical alternatives. A first direction
is to consider the dark energy paradigm, which, at least at an
effective level, can originate from various fields, such as a
canonical scalar field (quintessence) \cite{quint}, a phantom
field, that is a scalar field with a negative sign of the kinetic
term \cite{phant}, or the combination of quintessence and phantom
in a unified model named quintom \cite{quintom}. A second
direction is to modify gravity itself, using a function $f(R)$ of
the curvature scalar \cite{Sotiriou:2008rp}, higher derivatives in
the action \cite{Nojiri:2005jg}, braneworld extensions
\cite{brane}, string-inspired constructions \cite{string},
holographic properties \cite{holoext}, UV modifications
\cite{Horawa} etc.

Recently, a new alternative approach appeared in the literature
\cite{Ferraro:2006jd,Bengochea:2008gz,Linder:2010py}. It is based
on the old idea of the  ``teleparallel'' equivalent of General
Relativity (TEGR) \cite{ein28,Hayashi79}, which, instead of using
the curvature defined via the Levi-Civita connection, uses the
Weitzenb{\"o}ck connection that has no curvature but only torsion.
The dynamical objects in such a framework are the four linearly
independent vierbeins (these are \emph{parallel} vector fields
which is what is implied by the appellations ``teleparallel'' or
``absolute parallelism''). The advantage of this framework is that
the torsion tensor is formed solely from products of first
derivatives of the tetrad.  As described in \cite{Hayashi79}, the
Lagrangian density, $T$, can then be constructed from this torsion
tensor under the assumptions of invariance under general
coordinate transformations, global Lorentz transformations, and
the parity operation, along with requiring the Lagrangian density
to be second order in the torsion tensor. However, instead of
using the torsion scalar $T$ the authors of
\cite{Bengochea:2008gz,Linder:2010py} generalized the above
formalism to a modified $f(T)$ version, thus making the Lagrangian
density a function of $T$, similar to the well-known extension of
$f(R)$ Einstein-Hilbert action.

In comparison with $f(R)$ gravity, whose fourth-order equations
may lead to pathologies, $f(T)$ gravity has the significant
advantage of possessing second-order field equations. This feature
has led to a rapidly increasing interest in the literature, and
apart from obtaining acceleration
\cite{Bengochea:2008gz,Linder:2010py} one can reconstruct a
variety of cosmological evolutions
\cite{Myrzakulov:2010vz,Dent:2010bp}, add a scalar field
\cite{Yerzhanov:2010vu}, use observational data in order to
constrain the model parameters \cite{Wu:2010xk} and examine the
dynamical behavior of the scenario
 \cite{Wu:2010mn}.

All the previous investigations on $f(T)$ gravity have focused on
the background evolution. However, in order to reveal the full
scope and physical implications of the theory, one must delve into
the perturbative framework. Thus, one must first investigate the
perturbations of the pure gravitational sector, and extract the
corresponding dispersion relations, since such an analysis is
necessary in order to determine the stability (or lack thereof) of
the theory. Additionally, one should examine the complete set of
gravitational plus matter perturbations, since they are related to
the structure growth, and therefore their evolution could
constrain $f(T)$ gravity via comparison with the wealth of
precision data from observations. Both of the above investigations
are the main goals of this work. In particular, the present investigation is
interested in the first-order perturbations of a
Friedmann-Robertson-Walker universe under $f(T)$ gravity.

The plan of the work is as follows. In section \ref{model} we
present the cosmological scenario of $f(T)$ gravity at the
background level. In section \ref{vacuumperturbations} we examine
the first order perturbed equations for the scalar sector, we
extract the corresponding dispersion relations, we derive the
evolution equation for matter overdensities, and we evolve
perturbations for a specific model as an example. In section
\ref{vector} we extend our analysis in the vector and tensor
perturbations at linear order. Finally, section \ref{conclusions}
is devoted to the summary of our results.  The calculations in the
preceding sections are performed in the Newtonian gauge, however
for completeness we include Appendix A with the perturbed
equations for the scalar modes in the synchronous gauge.

\section{The cosmological background in $f(T)$ gravity}
\label{model}

In this section we study the cosmology of a  universe governed by
$f(T)$ gravity.  In this manuscript our notation is as follows:
Greek indices $\mu, \nu,$... run over all coordinate space-time 0,
1, 2, 3, lower case Latin indices (from the middle of the
alphabet) $i, j, ...$  run over spatial coordinates 1, 2, 3,
capital Latin indices $A, B, $... run over the tangent space-time
0, 1, 2, 3, and lower case Latin indices (from the beginning of
the alphabet) $a,b, $... will run over the tangent space spatial
coordinates 1, 2, 3.

As we stated in the Introduction, the dynamical variable of the
old ``teleparallel'' gravity, as well as its $f(T)$ extension, is
the vierbein field ${\mathbf{e}_A(x^\mu)}$. This forms an
orthonormal basis for the tangent space at each point $x^\mu$ of
the manifold, that is $\mathbf{e} _A\cdot\mathbf{e}_B=\eta_{AB}$,
where $\eta_{AB}=diag (1,-1,-1,-1)$. Furthermore, the
  vector $\mathbf{e}_A$ can be analyzed with the use of its components $e_A^\mu$
 in a coordinate basis, that is
$\mathbf{e}_A=e^\mu_A\partial_\mu $.

In such a construction, the metric tensor is obtained from the
dual vierbein as
\begin{equation}
\label{metrdef} g_{\mu\nu}(x)=\eta_{AB}\, e^A_\mu (x)\, e^B_\nu
(x).
\end{equation}
Contrary to General Relativity, which uses the torsionless
Levi-Civita connection, in the present formalism ones uses the
curvatureless Weitzenb\"{o}ck connection \cite{Weitzenb23}, $\overset{\mathbf{w}}{\Gamma}^\lambda_{\nu\mu} \equiv e_A{}^{\lambda}\partial_{\nu}e_{\mu}{}^A$, whose
 torsion tensor reads
\begin{equation}
 \label{torsion2}
{T}^\lambda_{\:\mu\nu}=\overset{\mathbf{w}}{\Gamma}^\lambda_{\nu\mu}-\overset
{\mathbf{w}}{\Gamma}^\lambda_{\mu\nu}=e^\lambda_A\:(\partial_\mu
e^A_\nu-\partial_\nu e^A_\mu).
\end{equation}
Moreover, the contorsion tensor, which equals the difference
between the Weitzenb\"{o}ck and Levi-Civita connections, is defined as
\begin{equation}
 \label{cotorsion}
K^{\mu\nu}_{\:\:\:\:\rho}=-\frac{1}{2}\Big(T^{\mu\nu}_{\:\:\:\:\rho}
-T^{\nu\mu}_{\:\:\:\:\rho}-T_{\rho}^{\:\:\:\:\mu\nu}\Big).
\end{equation}
Finally, it proves useful to define
\begin{eqnarray}
 \label{Stensor}
S_\rho^{\:\:\:\mu\nu}=\frac{1}{2}\Big(K^{\mu\nu}_{\:\:\:\:\rho}+\delta^\mu_%
\rho \:T^{\alpha\nu}_{\:\:\:\:\alpha}-\delta^\nu_\rho\:
T^{\alpha\mu}_{\:\:\:\:\alpha}\Big).
\end{eqnarray}
Note the antisymmetric relations $ T^{\lambda}{}_{\mu\nu} = - T^{\lambda}{}_{\nu\mu}$ and
$S_{\rho}{}^{\mu\nu} = -S_{\rho}{}^{\nu\mu} $, as can be easily
verified. Using these quantities
 one can define the so called ``teleparallel Lagrangian'' as
\cite{Hayashi79,Maluf:1994ji,Arcos:2005ec}
\begin{equation}  \label{telelag}
L_T\equiv S_\rho^{\:\:\:\mu\nu}\:T^\rho_{\:\:\:\mu\nu}.
\end{equation}
In summary, in the present formalism, all the information
concerning the gravitational field is included in the torsion
tensor ${T}^\lambda_{\:\mu\nu}$, and the teleparallel Lagrangian
$L_T$ gives rise to the dynamical equations for the vierbein,
which imply the Einstein equations for the metric.

From the above discussion one can deduce that the teleparallel
Lagrangian arises from the torsion tensor, similar to the way the
curvature scalar arises from the curvature (Riemann) tensor. Thus,
one can simplify the notation by replacing the symbol $L_T$ by the
symbol $T$, which is the torsion scalar \cite{Linder:2010py}.

While in teleparallel gravity the action is constructed by the
teleparallel Lagrangian $L_T = T$, the idea of $f(T)$ gravity is
to generalize $T$ to a function $T+f(T)$, which is similar in
spirit to the generalization of the Ricci scalar $R$ in the
Einstein-Hilbert action to a function $f(R)$. In particular, the
action in a universe governed by $f(T)$ gravity reads:
\begin{eqnarray}
\label{fTaction}
I = \frac{1}{16\pi G}\int d^4x e \left[T+f(T)+L_m\right],
\end{eqnarray}
where $e = \textrm{det}(e_{\mu}^A) = \sqrt{-g}$ and $L_m$ stands
for the matter Lagrangian. Variation of the action with respect to
the vierbein gives the equations of motion \begin{widetext}
\begin{eqnarray}\label{eom}
e^{-1}\partial_{\mu}(eS_{A}{}^{\mu\nu})[1+f'({T})]
-e_{A}^{\lambda}T^{\rho}{}_{\mu\lambda}S_{\rho}{}^{\nu\mu}[1+f'({T})] +
S_{A}{}^{\mu\nu}\partial_{\mu}({T})f''({T})-\frac{1}{4}e_{A}^{\nu}[T+f({T})]
= 4\pi Ge_{A}^{\rho}\overset {\mathbf{em}}T_{\rho}{}^{\nu},
\end{eqnarray}
\end{widetext} where a prime denotes the derivative with respect to
$T$ and the mixed indices are used as in $S_A{}^{\mu\nu} =
e_A^{\rho}S_{\rho}{}^{\mu\nu}$. Note that the tensor $\overset
{\mathbf{em}}T_{\rho}{}^{\nu}$ on the right-hand side is the usual
energy-momentum tensor, in which we added an overset label in order to
avoid confusion with the torsion tensor.

If we assume the background to be a perfect fluid, then the energy momentum tensor takes the
form
\begin{eqnarray}
\overset{\mathbf{em}}T_{\mu\nu} = pg_{\mu\nu} - (\rho +
p)u_{\mu}u_{\nu},
\end{eqnarray}
where $u^{\mu}$ is the fluid four-velocity. Note that we are
following the conventions of \cite{Weinberg:2008}, but with an
opposite signature metric. In the following sections we will be
interested in the perturbed energy-momentum tensor, and the signs
of the perturbed terms are not affected due to the indices mixing
(one upper and one lower).

Let us now focus on cosmological scenarios in a universe governed
by $f(T)$ gravity. Thus, throughout the work we consider the
common choice for the form of the vierbien, namely
 \be \label{weproudlyuse} e_{\mu}^A={\rm
diag}(1,a,a,a),
 \ee
which leads to a flat Friedmann-Robertson-Walker (FRW) background
geometry with metric
\begin{equation}
ds^2= dt^2-a^2(t)\,\delta_{ij} dx^i dx^j,
\end{equation}
where $a(t)$ is the scale factor. It has recently been shown \cite{Barrow:2010} 
that $f(T)$ gravity does not preserve local Lorentz invariance.  In principle then
one should study the cosmology resulting from a more general vierbein ansatz.  In 
this paper we specialize our attention to the diagonal form, leaving the more
general case for future investigation.

Using this vierbein, together with the
above fluid description of matter, one sees that equations
(\ref{eom}) lead to the background (Friedmann) equations
\begin{eqnarray}\label{background1}
&&H^2 = \frac{8\pi G}{3}\rho_m -\frac{f({T})}{6}-2f'({T})H^2\\\label{background2}
&&\dot{H}=-\frac{4\pi G(\rho_m+p_m)}{1+f'(T)-12H^2f''(T)}.
\end{eqnarray}
 In
these expressions we have introduced the Hubble parameter
$H\equiv\dot{a}/a$, where a dot denotes a derivative with respect
to coordinate time $t$. Moreover, $\rho_m$ and $p_m$ stand respectively for the
energy density and pressure of the matter content of the universe,
with equation-of-state parameter $w_m=p_m/\rho_m$. Finally, we
have employed the very useful relation
\begin{eqnarray}
T=-6H^2, \label{TH2}
\end{eqnarray}
which straightforwardly arises from evaluation of (\ref{telelag})
for the unperturbed vierbien (\ref{weproudlyuse}). Lastly, note
that General Relativity is recovered by setting $f(T)$ to a
constant (which will play the role of a cosmological constant), as
expected.

\section{Scalar Perturbations in $f(T)$ gravity}
\label{vacuumperturbations}

One of the most decisive tests for the reliability of a
gravitational theory is the examination of its perturbations. This
investigation reveals some of the deep features of the theory,
determining its stability and the growth of matter overdensities.
In this section we analyze in detail the linear-order scalar
perturbations of $f(T)$ gravity, leaving the vector and tensor
ones for the next section. In particular, we extract the full set
of gravitational and energy-momentum tensor perturbations in
subsection \ref{mattervierbpert}, and in subsection \ref{Stability} we
examine the stability of the theory. Finally, in subsection
\ref{growthpert} we examine the growth of of matter overdensities.
For simplicity we perform the calculations in the Newtonian gauge (for
completeness, Appendix A displays the scalar perturbations in the
synchronous gauge).

\subsection{Matter and scalar vierbein perturbations up to linear order}
\label{mattervierbpert}

We are interested in examining how the scalar vierbein
perturbations affect the equations of motion. We mention that
while in General Relativity the fundamental object is the metric,
which is then perturbed, in $f(T)$ gravity, as well as in the old
``teleparallel'' equivalent of General Relativity, the fundamental
object is the vierbein. Thus, the starting point is the vierbein
perturbation, which will then give rise to the perturbed metric.

Using the symbol  ${e}_{\mu}^A$ for the perturbed vierbein and
$\bar{e}_{\mu}^A$ for the unperturbed one, the scalar perturbation
writes
\begin{eqnarray}
e_{\mu}^A = \bar{e}_{\mu}^A + t_{\mu}^A,
\end{eqnarray}
where
\begin{eqnarray}
\label{pert1} &&\bar{e}_{\mu}^0 = \delta_{\mu}^0\,\,\,\,\,
\bar{e}_{\mu}^a = \delta_{\mu}^aa\,\,\,\,\,
\bar{e}^{\mu}_0 = \delta^{\mu}_0 \,\,\,\,\, \bar{e}^{\mu}_a = \frac{\delta^{\mu}_a}{a}\\
&&t_{\mu}^0 = \delta_{\mu}^0\psi \,\,\,\,\, t_{\mu}^a
=-\delta_{\mu}^a a\phi \,\,\,\,\, t^{\mu}_0 =
-\delta_{0}^{\mu}\psi \,\,\,\,\, t^{\mu}_a =
\frac{\delta^{\mu}_a}{a}\phi, \ \ \ \ \label{pert2}
\end{eqnarray}
with indicial notation as stated at the beginning of section
\ref{model}. As usual, we have kept terms up to first order in the
perturbations. Moreover, unless otherwise indicated, subscripts
zero and one will generally denote respectively zeroth and linear
order values of quantities.

In the above perturbations we have introduced the scalar modes
$\phi$ and $\psi$, which are functions of $\bx$ and $t$. These
symbols, as well as the various coefficients have been
conveniently chosen in order for the vierbein perturbation to
induce a metric perturbation of the known form in Newtonian gauge
with signature $(+---)$, namely
\begin{eqnarray}
\label{pertmetric}
 ds^2 = (1 + 2\psi)dt^2
-a^2(1-2\phi)\delta_{ij}dx^idx^j.
\end{eqnarray}

In the following, we make the simplifying assumption that the
scalar perturbations $t_{\mu}^A$ are diagonal, a procedure which is
sufficient to extract qualitative results about the stability of
the theory. Thus, the determinant becomes
\begin{eqnarray}
e = \textrm{det}(e_{\mu}^A) = a^3(1+\psi - 3\phi).
\end{eqnarray}

Proceeding forward, we calculate $T^{\lambda}{}_{\mu\nu}$ and
$S_{\lambda}{}^{\mu\nu}$ to first order under the perturbations
(\ref{pert1}) and (\ref{pert2}). The torsion reads
\begin{eqnarray}
\label{Tpert}
 T^{\lambda}{}_{\mu\nu} = (\bar{e}^{\lambda}_A +
t^{\lambda}_A)[\partial_{\mu}(\bar{e}_{\nu}^A + t_{\nu}^A) -
\partial_{\nu}(\bar{e}_{\mu}^A + t_{\mu}^A)],
\end{eqnarray}
and thus the second is easily calculable using (\ref{Stensor}).
After some algebra we find (indices are not summed over)
\begin{eqnarray}
&&T^{0}{}_{\mu\nu} = \partial_{\mu}\psi\delta_{\nu}^0 -
\partial_{\nu}\psi\delta_{\mu}^0\ \ \,;\ \ \, T^{i}{}_{0i} = H -
\dot{\phi} \nonumber\\
&&S_{0}{}^{0i} = \frac{\partial_i \phi}{a^2}\ \ \,;\ \ \,
S_{i}{}^{0i} = -H + \dot{\phi} + 2H\psi \nonumber\\
&& T^{i}{}_{ij} = \partial_j\phi \ \ \, \,;\ \ \, S_{i}{}^{ij} =
\frac{1}{2a^2}\partial_j (\phi - \psi).
\end{eqnarray}
Additionally, up to first order the torsion scalar defined in
(\ref{telelag}) is found to be
\begin{eqnarray}
{T}\equiv S_{\rho}{}^{\mu\nu}T^{\rho}{}_{\mu\nu} =T_0+T_1,
\end{eqnarray}
where
\begin{eqnarray}
 T_0=-6H^2
\end{eqnarray}
is the previously seen zeroth order result and
\begin{eqnarray}
\label{T1}
 T_1=12H(H\psi + \dot{\phi})
\end{eqnarray}
is the first order one. Thus, we can easily express $f(T)$ and its
derivatives up to first order as:
\begin{align}
&f(T)=f(T_0)+T_1\frac{df(T)}{dT}\Big|_{T=T_0}\equiv f_0+f_1\nonumber\\
&f'(T)=\frac{df(T)}{dT}\Big|_{T=T_0}+T_1\frac{d^2f(T)}{dT^2}\Big|_{T=T_0}\equiv f'_0+f'_1\nonumber\\
&f''(T)=\frac{d^2f(T)}{dT^2}\Big|_{T=T_0}+T_1\frac{d^3f(T)}{dT^3}\Big|_{T=T_0}\equiv
f''_0+f''_1,\ \ \ \ \ \ \label{FTexpansion}
\end{align}
that is the right hand sides of these equations are functions of
$T_0$ and linear functions of $T_1$.

Let us now consider the perturbations of the energy-momentum
tensor.
 The perturbations are then expressed as
\begin{eqnarray}
\label{T00pert}
\delta \overset {\mathbf{em}}T_0{}^0 &=& -\delta\rho_m\\
\delta \overset {\mathbf{em}}T_0{}^i &=& a^{-2}(\rho_m + p_m)(-\partial_i \delta u)\\
\delta\overset {\mathbf{em}} T_i{}^0 &=& (\rho_m +
p_m)(\partial_i\delta u) \label{Ta0pert}
\\
\delta\overset {\mathbf{em}} T_i{}^j &=& \delta_{ij}\delta p_m+\partial_i\partial_j\pi^{S}. \label{Tabpert}
\end{eqnarray}
where $\pi^S$ is the scalar component of the anisotropic stress.
Inserting everything in  (\ref{eom}) we finally obtain
\begin{eqnarray}
\label{E00}
 E_{0}^0 \equiv && (1+f'_0)(\nabla^2\phi) +
6(1+f'_0)H\dot{\phi} \ \ \ \ \ \nonumber
\\\nonumber
 &&\ \ \  + 6(1+f'_0)H^2\psi -3f'_1H^2 \ \  \ \ \
\\
 &&\ \ \  -\frac{T_1 +f_1}{4}=-4\pi G\delta \rho_m, \\
E_{0}^i\equiv &&(1+f'_0)\partial_i \dot{\phi} +
(1+f'_0)H\partial_i \psi \nonumber\\
   \label{E0a}
  &&
 -12H\dot{H}f''_0\partial_i \phi = -4\pi G(\rho_m + p_m)\partial_i \delta u,\\
 E_{a}^0 \equiv &&12H^2\partial_i\delta_a^i(\dot{\phi}+H\psi)f''_0-(1+f'_0)\partial_i\delta_a^i(\dot{\phi} +
H\psi)   \nonumber\\
 \label{Ea0} &&
  = 4\pi G
 (\rho_m + p_m)\partial_i\delta_a^i \delta u,
 \label{Eaa}
 \end{eqnarray}
 \begin{eqnarray} \label{Eaa}
E_{a}^i\delta_i^a\equiv &&\frac{f'_1}{a}\left(-3H^2-\dot{H}\right)
+ \frac{f''_1}{a}\left(12H^2\dot{H}\right)\ \ \ \ \ \ \ \ \ \ \ \
\ \ \ \ \ \nonumber
\\
 &&-\frac{(1+f'_0)}{2a}\sum_{b\neq a}\partial^j\delta_j^b\partial_i\delta^i_b(\psi-\phi) \nonumber \\
  &&-\frac{\phi (T_0 +f_0)}{4a}-\frac{T_1
 +f_1}{4a}
 \nonumber\\
  &&+ \frac{(1+f'_0)}{a}[6H\dot{\phi} + 6H^2\psi-3H^2\phi\nonumber \\
  &&
\ \ \ \ \ \ \ \ \ \ \ \ \ \ \ \ \ \ \ +\ddot{\phi}
+ \dot{H}(2\psi -\phi) + H\dot{\psi}]  \ \ \ \ \   \nonumber \\
  &&+\frac{f_0''}{a}(-24H\dot{H}\dot{\phi}-48\psi H^2\dot{H}-12H^2\ddot{\phi}\nonumber \\
  &&
\ \ \ \ \ \ \ \,  -12H^3\dot{\psi} + 12H^2\dot{H}\phi)\nonumber \\
  &&
\ \ \ \ \ \ \ \ \ \ \ \ \    = \frac{4\pi G}{a}(p_m\phi + \delta p_m),\\
E_{b;\,b\neq a}^i\delta_i^a\equiv
&&\frac{(1+f'_0)}{2}\partial_j\delta_b^j\partial^i\delta_i^a(\phi-\psi) \nonumber \\
&&=
4\pi G a^2\partial_j\delta_b^j\partial^i\delta_i^a\pi^{S}\label{Eab2}
\end{eqnarray}
where we have used the definition $\nabla^2 = \sum_i
\partial_i\partial^i$ and indices are summed over only when
explicitly shown with the $\sum$ symbol.

The above equations are perfectly general. In what follows, we set
the scalar anisotropic stress $\pi^S$ to zero for simplicity, a
choice which precludes the possibility of anisotropic expansion of
the Universe.  However, the cosmological consequences of dark
energy models with anisotropic stress have been previously
considered in the literature \cite{Mota}, and a detailed study of
the effects of anisotropic stress in $f(T)$ cosmology would be an
interesting avenue for future research.

The zero-anisotropic-stress assumption, which according to
\eqref{Eab2} implies $\phi=\psi$, along with $\delta p_m=0$, may
lead the system of perturbation equations becoming overdetermined.
Once the vanishing of the anisotropic stress is implemented, we then have four equations determining the three
remaining perturbation variables $\delta\rho_m$, $\phi$ and $\delta u$.
However, in the limit $f''(T)\simeq0$, equations \eqref{E0a} and
\eqref{Ea0} become identical, removing the over-determination. We
therefore conclude that the requirement of no anisotropic stress
imposes another constraint on $f(T)$ models, namely that
$f''(T)\simeq0$. However, note that this requirement on $f(T)$ might be relaxed for more general choices of vierbien than \eqref{weproudlyuse}.

\subsection{Stability}
\label{Stability}

In the previous subsection we derived the linear equations of
motion for the perturbations. In this subsection we will use them
to examine the stability of $f(T)$ gravity and extract the
dispersion relation for the pure gravitational perturbations.
Since we are concerned with the stability of the theory, in  this
subsection we ignore the matter sector in the set of equations
(\ref{E00})-(\ref{Eab2}). This is because if the gravitational
sector itself is unstable, the matter content cannot cure the
instability.

Working
in Fourier space we introduce the mode-expansion of $\phi$ as
\begin{eqnarray}
 \phi(t,\bx)=\int \frac{d^3k}{(2\pi)^\frac{3}{2}}
~\tilde{\phi}_k(t)e^{i\bk\cdot\bx}. \label{phiexpansion}
\end{eqnarray}
Therefore, inserting this into (\ref{T1}), we obtain the modes of
the Fourier transformed $T_1$ as
\begin{eqnarray}
\label{T1k}
 \tilde{T}_{1k}=12H^2\tpk+12H\dot{\tilde{\phi}}_k,
\end{eqnarray}
while (\ref{FTexpansion}) gives
 the modes of the Fourier transformed $f_1$ as
\begin{eqnarray}
\label{f1k}
 \tilde{f}_{1k}=\tilde{T}_{1k}\
\frac{df(T)}{dT}\Big|_{T=T_0}=\tilde{T}_{1k} f'_0,
\end{eqnarray}
the modes of the Fourier transformed $f'_1$ as
\begin{eqnarray}
\label{f1kprime} \tilde{f}'_{1k}=\tilde{T}_{1k}\
\frac{d^2f(T)}{dT^2}\Big|_{T=T_0}=\tilde{T}_{1k} f''_0,
\end{eqnarray}
and the modes of the Fourier transformed $f''_1$ as
\begin{eqnarray}
\label{f1kdoubleprime} \tilde{f}''_{1k}=\tilde{T}_{1k}\
\frac{d^3f(T)}{dT^3}\Big|_{T=T_0}=\tilde{T}_{1k} f'''_0,
\end{eqnarray}
where obviously $f_0$, $f'_0$ and $f''_0$ do not depend on $\phi$.

Inserting equations
(\ref{phiexpansion})-(\ref{f1kdoubleprime}) into (\ref{Eaa}), we obtain
\begin{eqnarray}
\label{phiddk}
\ddot{\tilde{\phi}}_k+\Gamma \dot{\tilde{\phi}}_k+\omega^2
\tilde{\phi}_k=0,
\end{eqnarray}
where $\omega^2$ and $\Gamma$ are respectively given by
\begin{widetext}
\begin{eqnarray}
\label{omega2}
\omega^2=\frac{\frac{3H^2}{2}+\dot{H}-\frac{f_0}{4}+\dot{H}f'_0-\(36H^4-48H^2\dot{H}\)f''_0+144H^4\dot{H}f'''_0}{1+f'_0-12H^2f''_0}
\end{eqnarray}
\begin{eqnarray}
\Gamma=\frac{4H\[1+f'_0-\(12H^2+9\dot{H}\)f''_0+36H^2\dot{H}f'''_0\]}{1+f'_0-12H^2f''_0}.
\end{eqnarray}
\end{widetext}

Equation (\ref{phiddk}) allows one to study the stability of any
given model. In particular, a model for which $\omega^2$ is
negative will clearly be unstable. Note that as mentioned before, it
is sufficient to consider a scenario without any matter
content. In such a
pure gravitational case, the background equation
(\ref{background2}) (that is with $\rho_m=0$, $p_m=0$) leads to a
constant $H$. Therefore, equation (\ref{omega2}) reduces to the
simple form
 \be
  \label{simple}
\omega^2=\frac{\frac{3H^2}{2}-\frac{f_0}{4}-36H^4 f''_0}{1+f'_0-12H^2f''_0}.
\ee

One can use relation (\ref{simple}) in order to
determine if a specific $f(T)$ model is free of instabilities. For
instance, we can insert the power-law ansatz $f(T)=\alpha (-T)^n$,
with $\alpha=(6H_0^2)^{1-n}/(2n-1)$ in the absence of matter
($H_0$ is the present Hubble parameter), and the exponential
ansatz $f(T)=-\alpha T\,(1-e^{pT_0/T})$, with
$\alpha=1/[1-(1-2p)e^p]$ in the absence of matter,
 considered in \cite{Linder:2010py}, and then we can
straightforwardly determine the allowed values for the
ansatz-parameters numerically. It is easy to check that in both these cases
the stability condition is satisfied for the phenomenologically
relevant ranges of parameters, that is $0<n<1$ for the power-law
model and $0<p<1$ for the exponential model.

\subsection{Growth of perturbations}
\label{growthpert}

Having examined the stability of $f(T)$ gravity, in this
subsection we switch on the matter sector, and examine the
fluctuations about the FRW background in the presence of matter.
This is a crucial subject in every cosmological scenario, and thus
it is the starting point for a thorough investigation of the
cosmology in the $f(T)$ framework.

In order to study the growth of perturbations, we assume for
simplicity a matter-only universe, that is we impose $p_m=0=\delta
p_m$. As usual, we define the matter overdensity $\delta$ as
 \be\delta\equiv\frac{\delta\rho_m}{\rho_m}.\ee

We now write equation (\ref{E00}) as
\begin{align}
\label{poisson}
&3H\(1+f'_0-12H^2f''_0\)\dot{\tilde{\phi}}_k\nonumber\\
&+\[\(3H^2+k^2/a^2\)\(1+f'_0\)-36H^4f''_0\]\tilde{\phi}_k\nonumber\\
&+4\pi G\rho\tilde{\delta}_k=0.
\end{align}
Note that this is the relativistic version of the Poisson equation
in $f(T)$ gravity. In summary, equations (\ref{phiddk}) and
(\ref{poisson}) can be used in order to evolve $\delta$ for a
given $f(T)$ model, which will allow for a comparison to
observational data.

Before proceeding to the investigation of equation
(\ref{poisson}), let us make a comment concerning the relation to
General Relativity. As we observe, a very interesting feature that
emerges from the above analysis, both of the pure gravitational
sector of the previous subsection as well as of the matter
inclusive sector of the present subsection, is that the linear
perturbations in $f(T)$ gravity reduce to those of General
Relativity in the limit where $f(T)$ is constant. For example, if
$f(T)={\rm const.}=-2\Lambda$, where $\Lambda$ is the cosmological
constant, then one can immediately see that equations
(\ref{E00})-(\ref{Eab2}) reduce to the well-known first-order
equations of General Relativity
\cite{Misner:1974qy,Weinberg:2008}, and furthermore equations
(\ref{phiddk}) and (\ref{poisson}) reduce to the well-known
equations for the growth of perturbations in $\Lambda$CDM
cosmology (see e.g. \cite{DuttaDent}).

Obtaining General Relativity as a limiting behavior at the level
of both the background and the perturbations, is not a standard
feature of a gravitational theory. In particular, any covariant
modified gravitational theory contains extra degrees of freedom at
the perturbational level, unless this theory can be conformally
transformed back to the standard General Relativity. These degrees
of freedom may have no impact at the background level (under
certain symmetries), but the perturbative modes could in principle
leave their imprints at first order, even if
one imposes the General Relativity limit. This is the case, for example, for
 Ho\v{r}ava-Lifshitz gravity \cite{Horawa}
where, although the theory in the Infra-Red coincides with General
Relativity, at the perturbative level one obtains the known strong
coupling problem, even in the IR \cite{Horava2}.

However, as we see from the analysis of the present work, $f(T)$
gravity seems to coincide with General Relativity when $f(T)$ is
constant, both at the background level and at the first-order
perturbation level, leaving possible differences to arise at
second order or beyond. Such a behavior is seen also in $f(R)$
gravity, both in its metric and Palatini formulation
\cite{fRpert0,fRpert}, and it is a significant feature of $f(T)$
gravity and a robust self-consistency test.

 We now proceed to the
investigation of the physical implications of a non-trivial
$f(T)$-ansatz, studying the growth of the overdensity for a
specific model. We choose the power-law model suggested in
\cite{Linder:2010py}:
\begin{eqnarray}
\label{powerlaw} f({T}) = \alpha(-T)^n,
\end{eqnarray}
where $\alpha=\(6H_0^2\)^{(1-n)}\(1-\Omega_{m0}\)/(2n-1)$ and
$H_0$ and $\Omega_{m0}$ refer to the Hubble parameter and the
matter density parameter at present.
\begin{figure}[ht]
\begin{center}
    \epsfig{file=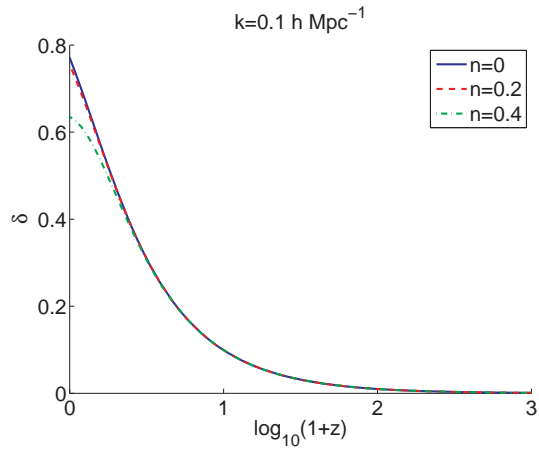,height=60mm}
    \caption
    {   \label{k0p1} \textit{The evolution  of the matter overdensity
    $\delta$ as a function of the redshift $z$, on a scale of $k=0.1h$ Mpc$^{-1}$, for three choices of $n$,
     for the power-law model given by (\ref{powerlaw}). }}
        \end{center}
\end{figure}

\begin{figure}[!]
\begin{center}
    \epsfig{file=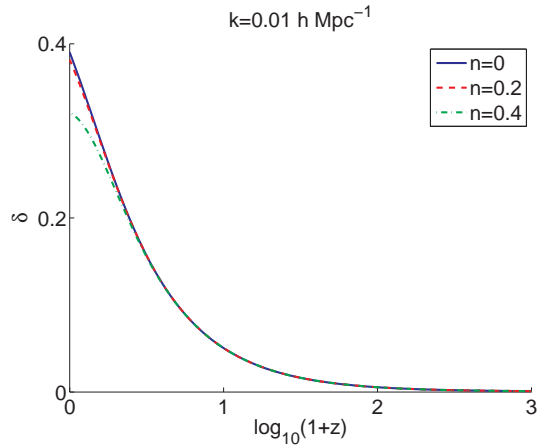,height=60mm}
    \caption
    {   \label{k0p01} \textit{The evolution of the matter overdensity $\delta$ as a function of the redshift $z$,
     on a scale of $k=0.01h$ Mpc$^{-1}$, for three choices of $n$, for the power-law model given by (\ref{powerlaw}).}}
            \end{center}
\end{figure}

\begin{figure}[!]
\begin{center}
        \epsfig{file=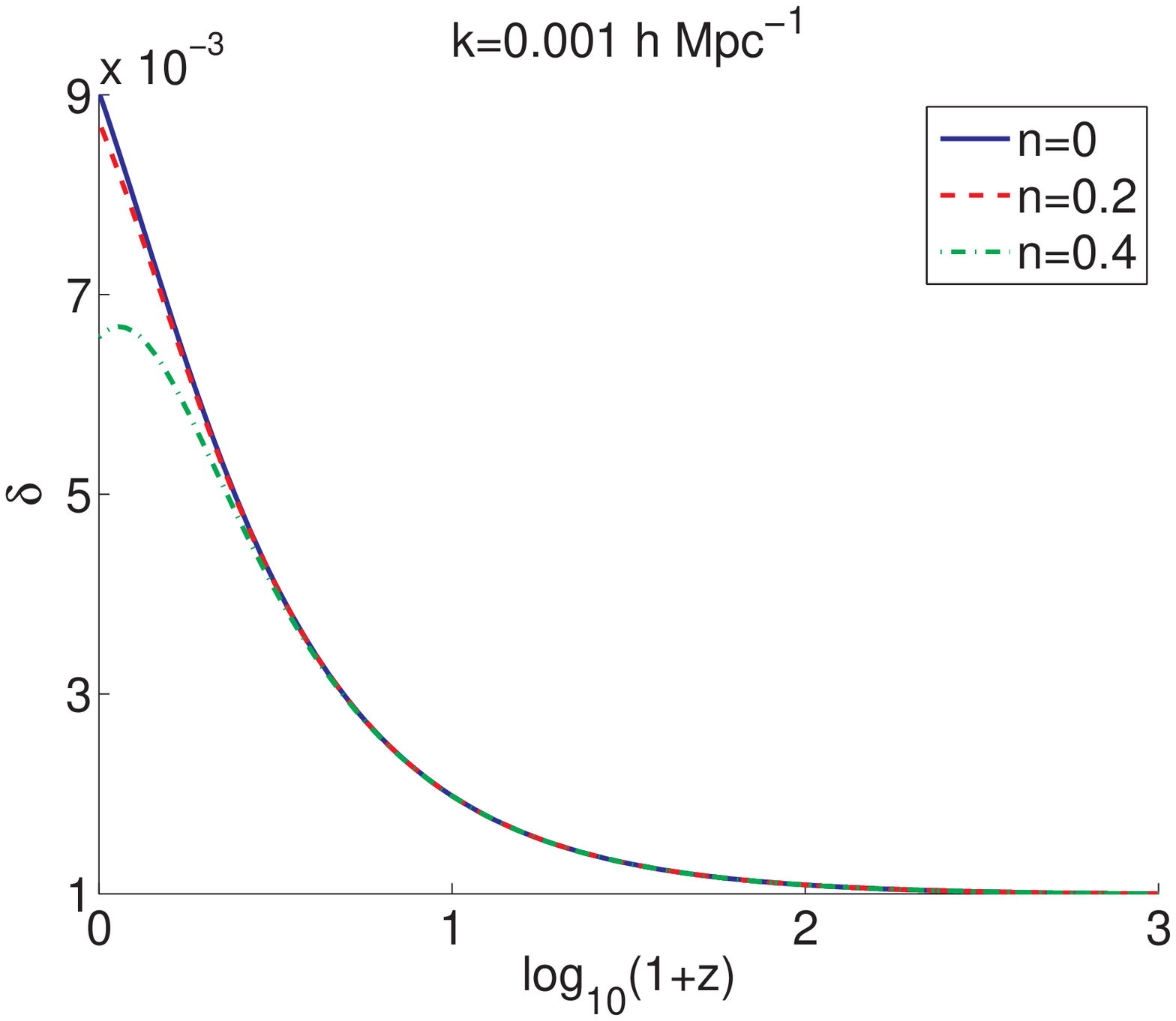,height=60mm}
    \caption
    {   \label{k0p001} \textit{The evolution of the matter overdensity $\delta$ as a function of the redshift $z$,
    on a scale of $k=0.001h$ Mpc$^{-1}$, for three choices of $n$, for the power-law model given by  (\ref{powerlaw}).}}
                \end{center}
\end{figure}

\begin{figure}[!]
\begin{center}
            \epsfig{file=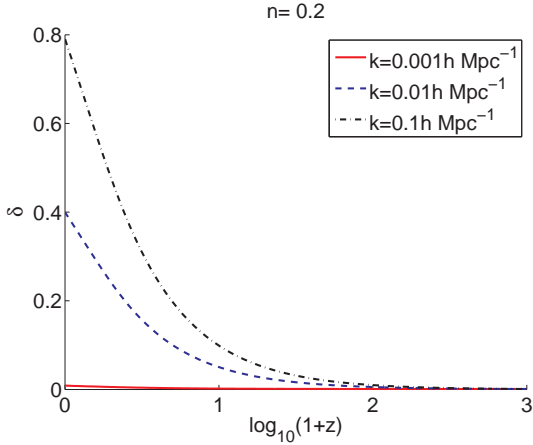,height=60mm}
    \caption
    {   \label{vark} \textit{The evolution of the matter overdensity $\delta$ as a function of the redshift $z$,
     for a fixed $n$ for the power-law model given by (\ref{powerlaw}), for three different scales.}}
    \end{center}
\end{figure}

Our results for the growth of perturbations, arising from a
numerical elaboration, are presented in Figures
\ref{k0p1}-\ref{k0p001}. In these figures we follow the growth of
the matter overdensity $\delta$, from the time of last scattering
to the present one, for different choices of $n$, for three
different $k$-scales. In Fig. \ref{vark} we depict the evolution
of $\delta$ for fixed $n=0.2$, but for three different scales. As
usual, we use the redshift $z$ as the independent variable,
defined as  $1 + z =a_0/a$ with $a_0$ the present scale-factor
value.

As expected, the $n=0$ case is identical to $\Lambda$CDM scenario
\cite{DuttaDent}. However, as $n$ increases we find that there is
a suppression of growth at smaller redshifts, which can act as a
clear distinguishing feature of these models. We also notice that
larger scales are more strongly affected than smaller ones.
Therefore, observations which target subhorizon scales close to
the horizon could, in principle, constrain these  models.

Finally, note that, as we mentioned in the end of section
\ref{mattervierbpert}, in order for our scenario to be free of
over-determination (since we have imposed the
zero-anisotropic-stress assumption which reduces the degrees of
freedom) we must have $f''(T)\simeq0$. This condition in the case
of the power-law ansatz requires $n\ll 1$, which is what we have
used for the numerical analysis. Interestingly enough, it is
exactly the same condition that is needed in order to acquire
an observationally compatible dark-energy and
Newton constant
phenomenology at the background level \cite{Linder:2010py} (for the power law ansatz
$n \simeq 1$ also gives $f''(T)\simeq0$ but we do not consider this case due to the
aforementioned phenomenological reason).

\section{Vector and Tensor  Perturbations in $f(T)$ gravity}\label{vector}

In the previous section we focused on the scalar perturbations of
$f(T)$ gravity, since they are sufficient to reveal the basic
features of the theory, allowing for a discussion of the growth of
matter overdensities. For completeness, in this section  we extend
our analysis in order to include the vector and tensor sectors of
the theory in the absence of matter.

The general perturbed vierbein at linear order reads as
\begin{eqnarray}
\label{fullpert}
e_{\mu }^{0} &=&\delta _{\mu }^{0}(1+\psi )+a\left( G_{i}+\partial
_{i}F\right) \delta _{\mu }^{i} \nonumber\\
e_{\mu }^{a} &=&a\delta _{\mu }^{a}(1-\phi )+a\left(
h_{i}{}^{a}+\partial _{i}\partial ^{a}B+\partial
_{i}C^{a}+\partial ^{a}C_{i}\right) \delta _{\mu
}^{i}\nonumber \\
e_{0}^{\mu } &=&\delta _{0}^{\mu }(1-\psi )\nonumber \\
e_{a}^{\mu } &=&\frac{1}{a}\left[\delta _{a}^{\mu }(1+\phi
)+\left( h^{i}{}_{a}+\partial ^{i}\partial _{a}B+\partial
^{i}C_{a}+\partial _{a}C^{i}\right) \delta _{i}^{\mu
}\right]\nonumber\\
&\ &-\left( G_{i}+\partial _{i}F\right) \delta _{a}^{i}\delta
_{0}^{\mu }.
\end{eqnarray}%
In these expressions, apart from the scalar modes $\phi$ and
$\psi$ of the previous section, we have introduced the transverse vector
modes $G_i$ and $C_i$, the transverse traceless tensor mode $h_i^a$, and the scalar
modes $F$ and $B$, the divergence of which will also contribute to
the vector sector. Similarly to the simple scalar case, the
coefficients on the above expressions have been chosen in order
for this vierbein perturbation to give rise to a perturbed metric
of the familiar form:

\begin{align}
&g_{00} =1+2\psi\nonumber \\
&g_{i0} =a\left[ \partial _{i}F+G_{i}\right] \nonumber\\
&g_{ij} =-a^{2}\left[ (1-2\phi )\delta _{ij}+h_{ij}+\partial
_{i}\partial _{j}B+\partial _{j}C_{i}+\partial _{i}C_{j}\right].
\end{align}

Let us make a comment here concerning the number of degrees of
freedom of the perturbed theory. As may be deduced straightaway,
$T^{\lambda}_{\mu\nu}$, $K^{\rho}_{\mu\nu}$ and $S^{\rho\mu\nu}$
are spacetime tensors under an infinitesimal coordinate
transformation of the form
 \be \label{gct} x^{\mu}\rightarrow
x^{\mu}+\epsilon^{\mu}.
 \ee
  This implies that the torsion scalar
$T$ is a generally covariant scalar, and thus actions of the form
of (\ref{fTaction}) are generally covariant as well as invariant under
 (\ref{gct}). As a result, for our choice of vierbien \eqref{weproudlyuse}, the number of degrees of freedom
(DOF) is identical to General Relativity. In particular, in $3+1$
spacetime dimensions, the metric, being symmetric, has 10
independent DOF. This is reflected in (\ref{fullpert}), which
comprises
\begin{itemize}
\item 4 scalar DOF $\psi$, $\phi$, $F$ and $B$ \item 4 vector DOF,
2 associated with each of the divergenceless vectors $G_i$ and
$C_i$ \item 2 tensor DOF associated with the transverse, traceless
and symmetric tensor $h_{ij}$
\end{itemize}
However, not all of these DOF are independent as there exist $3+1$
DOF associated with the coordinate transformation (\ref{gct}) (the
temporal part of $\epsilon^{\mu}$ is the scalar $\epsilon_{0}$,
and its spatial part can be decomposed into the gradient of a
scalar plus a divergenceless vector: $\partial_i \epsilon^S +
\epsilon_i^V$, leading to a total of 2 scalar and 2 vector DOF).
Subtracting these, we are left with a total of 6 DOF: 2 scalar, 2
vector, and 2 tensor, just as in the case of General Relativity.

We can therefore work in the Newtonian gauge, setting $F$ and $B$
to zero. This is easily understood since under (\ref{gct}), the
gauge transformation of $B$ is $-(2\epsilon^S)/a^2$, while that of
$F$ is $(1/a)(-\epsilon_0 -\dot{\epsilon}^S + 2H\epsilon^S)$.
Therefore, $\epsilon^S$ can be chosen in order to give rise to
$B=0$, and similarly an accompanying choice of $\epsilon_0$ will
lead to $F=0$.

From the above  analysis, it is clear that there are no extra
modes in $f(T)$ theories, which often show up in theories with
less symmetry than General Relativity. For example as discussed in
\cite{brandenberger}, in the case of Ho\v{r}ava gravity, which is
invariant under spatial diffeomorphisms and  space-independent
time re-parametrizations but not under space-dependent time
re-parametrizations, one can no longer choose the longitudinal
gauge, and thus an extra scalar DOF remains. However, note that extra DOF can still arise from more general choices of vierbien than \eqref{weproudlyuse}. The possible existence and phenomenology of these extra modes is beyond the scope of this work and warrants further study. 

Additionally, we choose a
gauge where the vector mode $C_i$ vanishes (through an appropriate choice of $\epsilon_i^{V}$).  As usual, the
vector modes are transverse, while the tensor mode
is transverse and traceless, namely
\begin{eqnarray}
\partial_i C^i =
\partial_i G^i = 0\,;\ \ \ \
\partial_ih^{ij} =
\delta^{ij} h_{ij} = 0.
\end{eqnarray}
Finally, we easily deduce the following relation between the
tensor perturbations in the vielbein and inverse vielbein:
\begin{eqnarray}
h_{\rho=1}{}^{a=1} = -h_{a=1}{}^{\rho=1}.
\end{eqnarray}

Using the above relations, the perturbed torsion tensor
(\ref{Tpert}) becomes
\begin{eqnarray}
T^0{}_{\mu\nu} &=& \partial_{\mu}\psi\delta^0_{\nu} - \partial_{\nu}\psi\delta_{\mu}^0 + a(\partial_{\mu}G_{\nu} -\partial_{\nu} G_{\mu})\nonumber\\
T^{i}{}_{0i} &=& H-\dot{\phi} + {\dot{h}_i{}^c}\delta_c^i\nonumber\\
T^{i}{}_{ij} &=& \partial_j\phi +
\partial_{i}h_{j}{}^c\delta_c^i-\partial_{j}h_{i}{}^c\delta_c^i,
\end{eqnarray}
where for notational compactness we have introduced a $G_0$ part of
$G_i$, which is zero. This torsion tensor inserted into
(\ref{Stensor}) leads to
\begin{eqnarray}
S_0{}^{0i} &=& \frac{1}{a^2}\partial_i\phi + \frac{1}{2a^2}\sum_j\partial_jh_{i}{}^a\delta_a^j + \frac{H}{a}G_i\nonumber\\
S_0{}^{ij} &=&\frac{1}{4a^3}\left(\partial_iG_j - \partial_jG_i\right)\nonumber\\
S_i{}^{0i} &=& -H + \dot{\phi} + 2H\psi + \frac{1}{2}{\dot{h}}_i{}^a\delta_a^i\nonumber\\
S_i{}^{j0} &=&
\frac{1}{4a}\left(\partial_iG_j-\partial_jG_i\right)\nonumber\\
\nonumber S_i{}^{ij} &=& \frac{1}{2a^2}\partial_j(\phi-\psi)
+\frac{1}{2a^2}\left(\partial_kh_j{}^a -
\partial_jh_k{}^a\right)\delta_a^k\\ &\ &\ + \frac{H}{a}G_j
+\frac{1}{2a}\dot{G}_j.
\end{eqnarray}
However, the torsion scalar is unaffected by the vector and tensor
modes, and thus it remains
\begin{eqnarray}
T\equiv T_0+T_1 = -6H^2 + 12H^2\psi + 12H\dot{\phi},
\end{eqnarray}
and likewise the determinant $e$ is still given by
\begin{eqnarray}
e = a^3(1+\psi-3\phi).
\end{eqnarray}

We now have all the necessary  machinery in order to extract the
equations of motion for the vector and tensor sectors. Following
the steps of the previous section, we can similarly decompose the
energy-momentum tensor into its vector and tensor components and
ignore the vector and tensor anisotropic stresses. We finally
obtain
\begin{eqnarray}
\label{vectoreqn0}
 [1+f'(T)]\nabla^2 G_j =0
\end{eqnarray}
for the vector mode. Since the quantity in square brackets is zero
only for the unphysical model $f(T)=-T$ (for which the action
(\ref{fTaction}) does not describe the gravitational sector
anymore), we can eliminate it, resulting in
\begin{eqnarray}
\label{vectoreqn} \nabla^2 G_j =0.
\end{eqnarray}
Therefore, the vector modes in $f(T)$ gravity decay as $1/a^2$,
that is similar to the General Relativity case.

For the tensor mode, we obtain
 {\small{
\begin{equation}\label{tensoreqn}
\bigg\{[1+f'(T)]\bigg(\frac{{\ddot{h}}_i{}^a}{2a}
-\frac{\nabla^2h_i{}^a}{2a}+\frac{3H{\dot{h}}_i{}^a}{2a}\bigg)
 - \frac{6H\dot{H}f''(T){\dot{h}}_i{}^a}{a}\bigg\}\delta_a^j = 0,
\end{equation}}}
Finally, similar to the scalar case, we can Taylor-expand the
derivatives of $f(T)$ using (\ref{FTexpansion}), and
Fourier-expand the vector and tensor modes, in order to obtain the
corresponding dispersion relations.  Moreover, we can split the
tensor sector into left-handed and right-handed polarizations.
However, such a detailed analysis of the gravitational wave
spectrum lies beyond the scope of this work. Here we retain only
the simple forms (\ref{vectoreqn}) and (\ref{tensoreqn}), since
they are adequate in order to reveal the basic features of the
behavior.

Concerning the tensor equation (\ref{tensoreqn}),
although there is a new friction term, there are no new mass terms, which
is a behavior similar to the scalar case of the previous section.
Therefore, we can safely conclude that, in general, $f(T)$
theories do not introduce massive gravitons; thus when $f(T)$
tends to a constant we do not obtain the typical problems of
massive gravity, which is a significant advantage of $f(T)$
gravity. Additionally, note that similar to the scalar case, in
the limit where $f(T)$ tends to a constant we do recover the
behavior of General Relativity at linear order, which is a
self-consistency test of the construction. Lastly, from the
equations of motion for scalar, vector and tensor perturbations
presented above, it is clear that these three classes of
perturbations decouple from one another in $f(T)$ gravity, just as
they do in case of General Relativity.

\section{Conclusions}
\label{conclusions}

In this work we investigated the recently developed $f(T)$
gravity, going beyond the simple background level. $f(T)$ gravity
is the extension of the ``teleparallel'' equivalent of General
Relativity, which uses the zero curvature Weitzenb\"{o}ck
connection instead of the  torsionless Levi-Civita connection, in
the same lines as $f(R)$ gravity is the extension of standard
General Relativity. In particular, we analyzed the first order
perturbations of $f(T)$ gravity.

Examining the scalar perturbations of $f(T)$ gravity in the
Newtonian gauge, we derived the perturbed equations of motion and
 extracted the corresponding dispersion relation. Therefore, in
constructing a realistic $f(T)$-gravitational scenario, one should
use an $f(T)$ ansatz that leads to non-negative $\omega^2$ in
(\ref{simple}), in order to obtain a theory free of instabilities.
Moreover, we showed that for the assumption of no scalar
anisotropic stress to be consistent, one needs the constraint
$f''(T)\simeq0$.

Additionally, we found that $f(T)$ gravity with $f(T)$ set to a
constant coincides with General Relativity, not only at the level
of the background but also for the first-order perturbations. This
is a significant advantage of the theory as compared to other
modified gravity paradigms.

Furthermore, as an example of an application of our formalism, we
followed the growth of perturbations in a specific $f(T)$ model,
namely the power-law ansatz proposed in \cite{Linder:2010py}. For
this model we found that on large subhorizon scales ($\mathcal{O}$(100 Mpc) or
larger), the evolution of the matter overdensity differs
from $\Lambda$CDM. Therefore, future precise observational data on
these scales could be used to constrain or rule out such models.

Finally, we investigated the vector and tensor perturbations at
linear order, extracting the corresponding equations of motion for
the vector and tensor modes. We showed that $f(T)$ gravity does
not introduce massive gravitons, which is a significant advantage.
Lastly, we verified again, as in the scalar sector, that in the
limit where $f(T)$ tends to a constant the theory tends to General
Relativity, both at the background as well as at the linear
perturbation level.

Clearly $f(T)$ cosmology presents a very rich behavior and
deserves further investigation.

\begin{acknowledgments}
The authors would like to thank Eric V. Linder for valuable
comments and Yifu-Cai, Franco Fiorini, Tomi Koivisto  and Gonzalo J. Olmo for
useful discussions.
\end{acknowledgments}

\appendix

\section{Scalar Gravitational and matter perturbations of $f(T)$ cosmology in the synchronous gauge} \label{synchronous}

Let us analyze the scalar gravitational and matter perturbations
in the synchronous gauge. The perturbed vierbein read
\begin{eqnarray}
e_{\mu}^i = \bar{e}_{\mu}^i + t_{\mu}^i,
\end{eqnarray}
with
\begin{eqnarray}
&&\bar{e}^0_{\mu} = \delta^0_{\mu}\,\,\,;\,\,\, \bar{e}^{a}_{\mu}
= \delta_{\mu}^a a\,\,\,;\,\,\,\bar{e}^{\mu}_0 = \delta^{\mu}_0
\,\,\,;\,\,\, \bar{e}^{\mu}_a  = \frac{\delta^{\mu}_a}{a}\nonumber\\
&&t_{\mu}^0 = 0\,\,\,;\,\,\, t_{\mu}^a = \delta_{\mu}^b
\frac{a}{2}\left(\delta_{b}^a A -a^2
\partial_b\partial^a B\right)\nonumber\\ &&
 t_{0}^{\mu} = 0\,\,\,;\,\,\,
t^{\mu}_a = -\delta_b^{\mu}\frac{1}{2a}\left(\delta_a^bA -
a^2\partial^b\partial_a B\right)\nonumber\\ && e = a^3\left(1  +
\frac{3}{2}A + \frac{1}{2}\sum_a \partial_i\partial_i B\right).
\end{eqnarray}
These symbols, as well as the various coefficients have been
conveniently chosen in order for the vierbein perturbation to
induce a metric perturbation of the known form in synchronous
gauge, namely
\begin{eqnarray}
ds^2 = dt^2 -a^2\left[\delta_{ij}(1 + A) + \partial_i\partial_j
B\right]dx^idx^j.
\end{eqnarray}
Inserting these into the perturbed torsion tensor (\ref{Tpert}),
and then to (\ref{Stensor}) we obtain
\begin{eqnarray}
T^{i}{}_{0\nu} &=& \delta^i_{\nu}\left(H + \frac{\dot{A}}{2}\right) -\frac{1}{2}\partial_t(a^2\partial^i\partial_bB)\delta^b_{\nu}\nonumber\\
T^i{}_{ji} &=& \frac{1}{2}\partial_j A\nonumber\\
S_0{}^{0i} &=& \frac{1}{2}\partial^i A\nonumber\\
S_i{}^{ij} &=& \frac{1}{4}\partial^j A\nonumber\\
S_i{}^{0j} &=& -\frac{1}{4}\partial_t(a^2\partial_i\partial^jB)\nonumber\\
S_{i}{}^{0i} &=& -\left(H+\frac{\dot{A}}{2}\right) +
\frac{1}{4}\partial_t(a^2\sum_{b\neq a}\partial^j\delta_j^b\partial_i\delta^i_bB)\nonumber\\
&=& -\left(H+\frac{\dot{A}}{2}\right) +
\frac{1}{4}\partial_t(a^2\nabla^2 B -a^2\partial_i\partial^i B).\
\
\end{eqnarray}
Furthermore, up to second order the torsion scalar defined in
(\ref{telelag}) is found to be
\begin{eqnarray}
{T} = -6H^2 -\sum_i\frac{1}{2}\partial^iA\partial_i A -6H\dot{A} +
2H\partial_t(a^2\nabla^2 B).\ \ \
\end{eqnarray}
Finally, concerning the first-order matter perturbations, we use
the results (\ref{T00pert})-(\ref{Tabpert}).

Inserting these results into (\ref{eom}) we extract the perturbed
equations of motion as:
\begin{widetext}
\begin{eqnarray}
\nonumber
E_0^0 &=& -\frac{(1 +f'_0)}{2}\nabla^2 A -3H^2f'_1({T}) - 3H\dot{A}(1+f'_0) + H\partial_t(a^2\nabla^2 B)(1+f'_0)-\frac{T_1 + f_1({T})}{4} \\
 &=& 4\pi G \delta \overset {\mathbf{em}}T_{0}{}^0,
\\
E^i_0 &=& \frac{1}{2}\partial^i\dot{A}(1+f'_0)
+H\partial^iA(1+f'_0)-6H\dot{H}f''_0({T})\partial^iA = 4\pi G
\delta \overset {\mathbf{em}}T_0{}^i,\\
 \nonumber
 E^0_a &=&
\frac{(1+f'_0)}{2a}\partial_i\delta_a^i\dot{A} +
\frac{Hf''_0({T})}{a}\partial_i\delta_a^i\left[-\frac{1}{2}\sum_b
\partial^j\delta_j^bA\partial_i\delta_b^iA -6H\dot{A} + 2H\partial_t(a^2\nabla^2
B)\right]\\&& = \frac{4\pi G}{a} \delta \overset
{\mathbf{em}}T_i{}^0\delta_a^i,\\\nonumber E^i_b &=&
-\frac{(1+f'_0)}{4a^3}\partial_t[a^2\partial_t(a^2\partial_j\delta_b^j\partial^iB)]+\frac{(1+f'_0)}{4a}\partial_j\delta_b^j\partial^iA
+ \frac{H(1+f'_0)}{4a}\partial_t(a^2\partial_j\delta_b^j\partial^iB)
-\frac{(1+f'_0)}{2a^3}\partial_t(Ha^4\partial^i\partial_j\delta_b^j B)\\&&+
3\frac{f''_0({T})}{a}H\dot{H}\partial_t(a^2\partial_j\delta_b^j\partial^iB)
+ 6aH^2\dot{H}f''_0(T)\partial^i\partial_j\delta_b^jB
-\frac{a}{8}\partial_j\delta_b^j\partial^i B (T_0 + f_0) = \frac{4\pi G}{a}
\delta \overset {\mathbf{em}}T_j{}^i\delta_b^j,\\\nonumber
 E_a^i\delta_i^a &=&
-\frac{f'_1}{a}\left[3H^2 + \dot{H}\right]
\\\nonumber &+&
(1+f'_0)\left\{\frac{1}{a^3}\left\{\partial_t(-Ha^2)\left(-\frac{3A}{2}+a^2\nabla^2
B\right)\right. \right.\\\nonumber &+&
\left.\partial_t\left\{a^2\left[-HA -\frac{\dot{A}}{2} +
\frac{1}{4}\partial_t(a^2\nabla^2 B -a^2\partial_i\partial^i B) +
\frac{H}{2}(a^2\nabla^2
B-a^2\partial_i\partial^iB)\right]\right\}\right\}\\\nonumber &-&
\frac{1}{4a}(\nabla^2 A -\partial_i\partial^iA) \\\nonumber
&-&\left. H\dot{A} +\frac{H}{4}\partial_t[a^2(\nabla^2 B +
\partial_i\partial^i B)]\right\}
 +\frac{f''_1({T})12H^2\dot{H}}{a}\\\nonumber
&+&\frac{f''_0({T})}{a}\left[-\frac{H}{2}\partial_t(\sum_a\partial_iA\partial^iA)
- 6H^2\ddot{A}-6H^2\dot{H}A\right.\\\nonumber  &-& \left.
H\dot{H}\partial_t(a^2\nabla^2 B)+2H^2\partial_t^2(a^2\nabla^2
B)+6H^2\dot{H}a^2\partial_i\partial^i B +
3H\dot{H}\partial_t(a^2\partial_i\partial^iB)\right]\\\nonumber
&-&\frac{(T_1 +f_1)}{4a} + (T_0 + f_0)\left(\frac{1}{8a}A
-\frac{a}{8}\partial_i\partial^i B\right) \\  &=& 4\pi G \delta
\overset {\mathbf{em}}T_1{}^1 -4\pi G p_m \frac{1}{2a}A + 4\pi
Gp_m \frac{a}{2}
\partial_i\partial^i B,
\end{eqnarray}
\end{widetext}
where we have used the definition $\nabla^2 = \sum_i
\partial_i\partial^i$, and indices are summed over only when
explicitly shown with the $\sum$ symbol.

\end{document}